# Kapitza resistance in basic chain models with isolated defects

Jithu Paul, O. V. Gendelman

*Faculty of Mechanical Engineering, Technion–Israel Institute of Technology, Haifa, Israel*

Kapitza thermal resistance is a common feature of material interfaces. It is defined as the ratio of the thermal drop at the interface to the heat flux flowing across the interface. One expects that this resistance will depend on the structure of the interface and on the temperature. We address the heat conduction in one-dimensional chain models with isotopic and/or coupling defects and explore the relationship between the interaction potentials and simulated properties of the Kapitza resistance. It is revealed that in linear models the Kapitza resistance is well-defined and size-independent (contrary to the bulk heat conduction coefficient), but depends on the parameters of thermostats used in the simulation. For β-FPU model one also encounters the dependence on the thermostats; in addition, the simulated boundary resistance strongly depends on the total system size. Finally, in the models characterized by convergent bulk heat conductivity (chain of rotators, Frenkel-Kontorova model) the boundary resistance is thermostat- and size-independent, as one expects. In linear chains, the Kapitza resistance is temperature-independent; thus, its temperature dependence allows one to judge on significance of the nonlinear interactions in the phonon scattering processes at the interface.

## 1. Introduction

Interfacial thermal resistance was first reported by Kapitza [1] after his seminal explorations on the superfluidity of liquid helium II. He observed the existence of temperature discontinuity at the interface of liquid Helium II and a solid at very low temperature. This achievement led to a development of general concept of boundary thermal resistance, broadly referred to as Kapitza resistance. This boundary resistance is an important factor governing the thermal transport, in addition to more common and better known thermal resistance of a bulk. In 1952, Khalatnikov developed acoustic mismatch model (AMM) [2], which fueled the theoretical understanding of Kapitza resistance based on the transmission coefficient of acoustic phonons. The Khalatnikov theory was first applied to solid-solid interface by Little in 1959 [3]. It should be mentioned that the analytical calculation for perfect solid-solid junction gives non-zero Kapitza resistance, which is a major drawback of Khalatnikov theory. Sluckin [4] calculated the energy transmission coefficient based on a simple hard particle model at the interface and the prediction was in considerable agreement with the experimental results. Then, the phonon transmission coefficient for one-dimensional harmonic chain was derived in [5]. The attempt to predict the Kapitza resistance with modification for heat current (correction for non-zero Kapitza resistance at perfect interface) in Khalatnikov theory can be found in [6] and the result is compared with molecular dynamics simulations. The AMM fails to give a realistic description with the experimental results at higher temperatures. Swartz developed diffuse mismatch model (DMM) [7] based on the assumption that diffuse scattering of phonons take place at the interface. The study claims that the AMM is more realistic at very low temperature, whereas the DMM is at high temperature. More detailed history of theoretical and experimental studies of Kapitza resistance can be found in the review articles [8-11].

It is well-known that the relationship between microstructure of solids and macroscopic heat transport properties remains elusive to large extent. This lack of knowledge, together with new technological possibilities for experimentation, led to substantial bulk of research on a topic of heat conduction in low-dimensional systems. Most impressive (albeit, still ongoing) progress has been achieved for one-dimensional systems. Analytical results based on stochastic calculations prove, in both mass disordered and ordered harmonic systems, thermal conductivity diverges with chain length $N$ [12,13]. These studies also provided analytical expressions [13] for the non-equilibrium temperature distribution for harmonic systems. For nonlinear systems, the progress required broad reliance on numeric methods. The famous work by Fermi-Pasta-Ulam (FPU) [14] was one of the important milestone in the history of numerical approach towards the heat conduction problem which demonstrated that mere nonlinearity of the interactions cannot guarantee the equipartition of energy between different modes. All one-dimensional models can be crudely divided into two broad universality classes – with divergent and convergent heat conduction coefficient in thermodynamical limit. The examples for the systems with divergent ($\kappa \propto N^\alpha$) heat conductivity are FPU chain [15, 16], diatomic Toda lattice [17] and gas of colliding particles [18-20]. Convergent heat conductivity is exhibited by chain of rotators (momentum conserving model) [21], linear chain with on-site potential (Frenkel-Kontorova model, linear chain with double well on-site potential, linear chain with hyperbolic sine on-site potential) [22, 23], non-zero sized hard disks with periodic piecewise linear on-site potential [24] and the class of models that are capable of dissociation in the chain [25].

In all studies devoted to the heat conductivity in one-dimensional systems, the major issue is the effect of the interaction nonlinearity on the heat transport. The studies mentioned above had clarified this effect for the problem of the bulk conductivity. The phenomenon of Kapitza boundary resistance received much less attention in this respect. One normally perceives that the Kapitza resistance depends on particular structure of the interface and on the temperature, but not on the size of the bulk at both sides, and not on peculiarities of thermostats used for the numeric simulation. The goal of present paper is to assess whether the Kapitza resistance simulated in one-dimensional models with different potentials of interaction conforms to this common perception. Instead of considering the interface between two bulk fragments we will simulate the heat flux through isolated defect (particle with different mass, or defective link) in otherwise homogeneous chain. This simplification delivers qualitatively similar results, but allows one to avoid complications related to possible non-reciprocity of the system [26-28]. General setting is described in Section 2. The linear chains are discussed in Section 3, and the nonlinear ones – in Section 4, followed by conclusions.

## 2. Description of the model and general settings

We are going to consider a one-dimensional chain with nearest neighbor interactions. All parameters are adopted as dimensionless. The Hamiltonian of the system can be represented as,

$$H = \sum_{n=1}^{N} \frac{p_n^2}{2m_n} + \sum_{n=1}^{N} U(x_{n+1} - x_n) + V(x_n) \tag{1}$$

Here $p_n$, $x_n$ and $m_n$ are the momentum, position and mass of the $n^{\text{th}}$ particle respectively. $U(x_{n+1} - x_n)$ is the potential energy of interaction between each neighboring particle pair and $V(x_n)$ is the on-site potential. We use fixed boundary conditions with $x_0 = x_{n+1} = 0$. In current paper, we are going to consider the following models:

Harmonic potential,
$$U(x) = \frac{1}{2}kx^2, V(x_n) = 0 \tag{2}$$

β-FPU potential:
$$U(x) = \frac{1}{2}kx^2 + \frac{\beta}{4}x^4, V(x_n) = 0 \tag{3}$$

Periodic potential (chain of rotators):
$$U(x) = 1 - \cos x, V(x_n) = 0 \tag{4}$$

Frenkel-Kontorova potential:
$$U(x) = \frac{1}{2}kx^2, V(x_n) = 1 - \cos x_n \tag{5}$$

Coefficient $k$ is the stiffness constant, and $\beta$ is the strength of the cubic non-linearity. In all the following sections, except for the defect mass $m$ or defective link with stiffness $k$, all other masses and stiffness constants are set to unity.

The ends of the chain are connected to the hot thermostat at the left end and to the cold thermostat at the right end. Terminal $N_\pm$ particles are attached to thermostats; the latter are maintained at temperatures $T_\pm$ (the signs represent hot and cold thermostats respectively). The pre-set thermostat temperatures (not necessarily coinciding with actual kinetic temperatures of the respective chain particles) can be represented as $T_\pm = T(1 \pm \Delta)$, where $\Delta = 0.1$ and $T$ is the average temperature of the system. The number of particles in the chain segment (between the thermostats) is denoted as $N$. We use Langevin dynamics to govern the thermostats. The dynamics of the system is described by the following equations [15, 23]:

$$m_n x_n'' = -\frac{\partial H}{\partial x_n} - \gamma x_n' + \xi_n^+, n = 1, \ldots N_+$$

$$m_n x_n'' = -\frac{\partial H}{\partial x_n}, n = N_+ + 1, \ldots N_+ + N - N_- \tag{6}$$

$$m_n x_n'' = -\frac{\partial H}{\partial x_n} - \gamma x_n' + \xi_n^-, n = N_+ + N + 1, \ldots N_+ + N + N_-$$

$x_n''$ and $x_n'$ are the acceleration and velocity of the $n^{\text{th}}$ particle, $\gamma$ is the damping coefficient and $\xi_n^\pm$ is the corresponding Gaussian white noise, which is normalized as follows,

$$\langle \xi_n^\pm(\tau_n) \rangle = \langle \xi_n^\pm(\tau_1)\xi_{n'}^\mp(\tau_2) \rangle = 0$$

$$\langle \xi_n^{\pm}(\tau_1)\xi_{n'}^{\pm}(\tau_2)\rangle = 2\gamma T_{\pm}\delta_{nn'}\delta(\tau_2 - \tau_1) \tag{7}$$

where $\tau_n$ is the different times in the simulation and $\tau$ represent total simulation time steps. We performed the time integration using the Verlet algorithm.

The average kinetic temperature of each particle is calculated as,

$$T_n = \langle t_n(\tau_n)\rangle_\tau = \lim_{\tau \to \infty} \frac{1}{\tau}\int_0^\tau m_n {x'_n}^2(\tau_n)d\tau_n \tag{8}$$

The instantaneous temperature is given as, $t_n(\tau_n) = m_n{x'_n}^2(\tau_n)$. The heat flux through the system is evaluated as follows:

$$J_n = \langle j_n(\tau_n)\rangle_\tau = \lim_{\tau \to \infty} \frac{1}{\tau}\int_0^\tau j_n(\tau_n)d\tau_n \tag{9}$$

Here $j_n$ is the instantaneous heat flux through the link at time $\tau_n$ and is given as [15],

$$j_n = -\frac{1}{2}(\dot{x}_n + \dot{x}_{n+1})F(x_{n+1} - x_n) \tag{10}$$

The simulation has been done in homemade FORTRAN-95 codes with a time step of 0.02. We have verified that the results are well-converged, even for larger time step. All the initial conditions for the system is taken as zero, $\{x_n(0) = 0, x'_n(0) = 0\}_{n=1}^N$. The system will be equilibrated for $10^6$ time steps with equal temperatures for hot and cold thermostats, then the production run is carried out for $10^9$ time steps.

The Kapitza resistance is defined as, $R_K = \frac{\Delta T}{J}$ [1]. Here $\Delta T$ is the temperature drop at the defect location and $J$ is the heat current through the chain. For linear chains, it is possible to evaluate the difference between average temperature of the left and right segments with respect to the defect. For the nonlinear chains, $\Delta T$ is measured by taking the temperature difference between the particle just before and after the defect, where the complete temperature drop occurs. In the following sections, the defect position will be always at the middle of the chain. Again, everywhere below we consider the Kapitza resistance caused by a single isolated defect placed at the middle of the chain, unless otherwise stated.

## 3. Linear chain
### 3.1. Numerical simulation results

For analysis of the linear systems, in addition to the straightforward molecular dynamics described above, we used also analytical-numerical approach based on RLL [12] method. The method is based on evaluation of the stationary state for the harmonic chain. The condition for the stationary state can be represented as:

$$a.b + b.a^T = d \tag{11}$$

Here, $a = \begin{bmatrix} 0 & -M^{-1} \\ K & M^{-1}G \end{bmatrix}$, $d = \begin{bmatrix} 0 & 0 \\ 0 & D \end{bmatrix}$ and $a^T$ is the transpose of matrix $a$. $M$ is the mass diagonal-matrix and $K$ is the stiffness coefficient tridiagonal-matrix. Matrix $G_{ij} = \gamma\delta_{ij}(\delta_{i,N_+} + \delta_{i,N_-})$, where $\gamma$ is the friction coefficient for the Langevin thermostat and $D_{ij} = 2\gamma\delta_{ij}(T_+\delta_{i,N_+} + T_-\delta_{i,N_-})$. Equation (11) is a type of Sylvester equation which can be solved for $b$ [29]. Here, $a$, $b$ and $d$ are $2N \times 2N$ matrices; and $M$, $G$ and $D$ are $N \times N$ matrices. Thus, the first step is to create the matrices $M$, $G$ and $D$ for the harmonic chain of $N$ particles whose $N_+$ and $N_-$ particles at the ends are attached to the hot ($T_+$) and cold ($T_-$) thermostats respectively with a coupling friction $\gamma$. The second task is to assemble matrices $M$, $G$ and $D$ to their parent matrices $a$ and $d$, and to solve Eq. (11) numerically for $b$. The numerically solved matrix $b$ will provide the steady state local temperature $T_n = b_{N+n,N+n}/m_n$ and local heat flux $J_n = k_n(b_{n+1,N+n} - b_{n,N+n})/m_n$ distributions.

In linear model, the interaction is described by potential function (2). We studied how the parameters of the defect effect the Kapitza resistance. First we studied the effect of defect strength on Kapitza resistance. We considered the strength of two types of defects, isotopic defect (Fig. 1(a)) and defective link (Fig. 1(b)). As evident from Fig. 1, the Kapitza resistance increases with the strength of the defect, either the defect is weaker or stronger compared to the rest of the system.

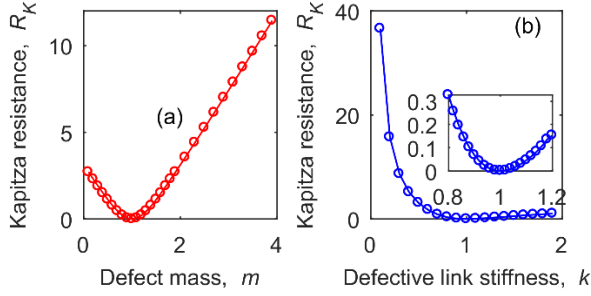

FIG. 1. The Kapitza resistance $R_K$ variation with the defect mass $m$ (a) and variation with the defective link stiffness $k$ (b). For (a) the parameters are $N = 101$, $T_+ = 1.1$, $T_- = 0.9$, $N_\pm = 1$, $\gamma = 1$ and for (b) all the parameters are same except $N = 100$. The defect is placed at the exact middle. The inset in (b) shows the high resolution plot for $0.8 \leq k \leq 1.2$. The line is to guide the eye.

It is well-known that the heat flux in a linear chain is proportional to the temperature difference between the chain extrema rather than to the temperature gradient [12] and the thermal conductivity diverges with the chain length as $\kappa(N) \propto N$. The simulation results show the temperature drop at isotopic defect location is also proportional to the heat flux and the Kapitza resistance does not depend on the chain length (Fig. 2). The chain lengths are taken in the interval $N = 21 - 2001$.

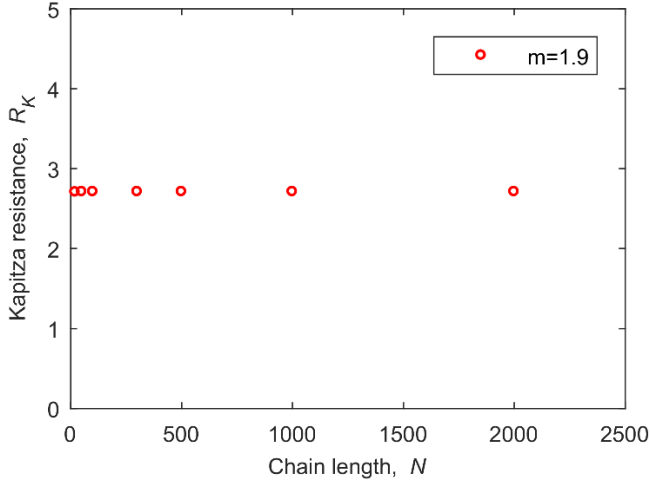

FIG. 2. Dependence of the Kapitza resistance $R_K$ on the chain length $N$ for the chain with linear interaction ($T_+ = 1.1$, $T_- = 0.9$, $N_\pm = 1$, $\gamma = 1$).

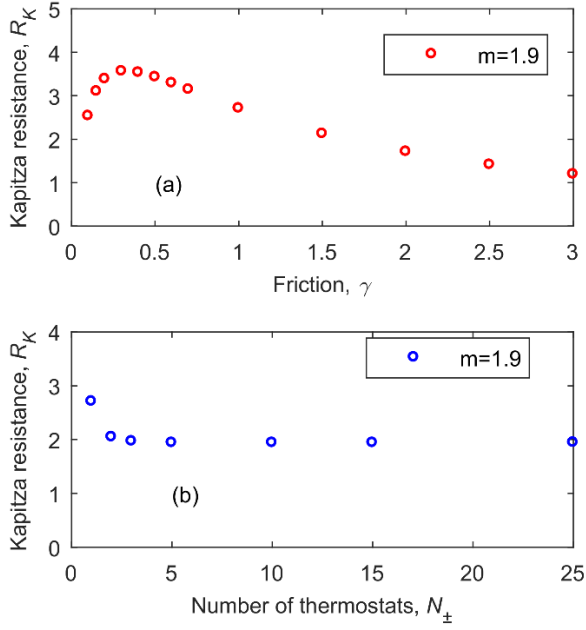

FIG. 3. Dependence of the Kapitza resistance $R_K$ on the coupling friction $\gamma$ (a) and on the number of thermostats $N_\pm$ (b) for the chain with linear interaction ($T_+ = 1.1$, $T_- = 0.9$, $N = 101$).

For a general mass-disordered linear chain, the available analytical results [13, 30] suggest that the heat flux and the local temperature in the chain depend on the thermostat characteristics. Since the linear chain with isotopic defect is a special case of mass disordered linear chain, one can expect that the Kapitza resistance also will depend on the thermostat characteristics. This is clearly observed in Fig. 3. The Kapitza resistance first increases and then decreases with coupling friction (see Fig. 3(a)). As for the number of the thermostats, the Kapitza resistance initially decreases and then remains constant (see Fig. 3(b)). The coupling friction and the number of thermostats are taken in the interval $\gamma = 0.1 - 3$ and $N_\pm = 1 - 25$ respectively. It is obvious that in a linear chain, the heat transport is take place by non-interacting phonons. As a result, the heat flux only scales with the chain average temperature and the ratio of $\Delta T$ and $J$ remains constant, that is the Kapitza resistance should be independent of temperature in linear system.

### 3.2. Approximate analytic evaluation of the Kapitza resistance.

We consider a homogeneous linear one-dimensional lattice with isolated defect, that occupies the sites with numbers $n_L \leq 0 \leq n_R$. We assume that the waves come from left side of the lattice, some of them pass through the defect and some are reflected. Under such assumptions, the field of displacements caused by a single monochromatic incident wave with unit amplitude, can be expressed as follows:

$$x_n(t) = \begin{cases} \exp(i(\omega t - qn)) + \alpha(\omega)\exp(i(\omega t + qn)), & n \leq n_L \\ \beta(\omega)\exp(i(\omega t - qn)), & n \geq n_R \end{cases} \quad (12)$$

Here it is assumed that the frequency belongs to the propagation zone of the lattice, $\alpha(\omega)$ and $\beta(\omega)$ are frequency-dependent reflection and transmission coefficients respectively. Due to energy conservation, they must obey well-known formula:

$$|\alpha(\omega)|^2 + |\beta(\omega)|^2 = 1 \quad (13)$$

For the defect lattice fragment $n_L \leq n \leq n_R$ the displacement field is determined by specific nature of the defect.

We also assume that the defect is sufficiently distant from the thermostats. Assuming that the incident heat flux has random phases, the energy densities (or, equivalently, the temperatures) of the lattice fragments leftwards and rightwards of the defect can be expressed as follows:

$$T_L = \int_{P_\omega}(1 + |\alpha(\omega)|^2)f(\omega)d\omega \, ; \, T_R = \int_{P_\omega}(|\beta(\omega)|^2)f(\omega)d\omega \quad (14)$$

Here $f(\omega)$ denotes the energy distribution in the incident heat flux, $P_\omega$ denotes the propagation zone of the lattice. The heat flux is expressed as follows:

$$J = \int_{P_\omega} (|\beta(\omega)|^2) f(\omega) v_{gr}(\omega) d\omega; \quad v_{gr}(\omega) = \frac{d\omega(q)}{dq} \tag{15}$$

Here $\omega(q)$ is the dispersion relation of the lattice. Thus, with account of (13-15), the expression for the Kapitza resistance associated with the considered defect will take the following shape:

$$R_K = \frac{T_L - T_R}{J} = \frac{2\int_{P_\omega} |\alpha(\omega)|^2 f(\omega) d\omega}{\int_{P_\omega} (|\beta(\omega)|^2) f(\omega) v_{gr}(\omega) d\omega} \tag{16}$$

As a specific example, we consider a simple linear chain with a single defective link. Namely, without reducing the generality, all masses and all coupling stiffnesses are set to unity, with exception of the coupling spring between masses $n = 0$ and $n = 1$. Stiffness of this coupling spring is set to $k > 0$. This model is described by the following equations of motion:

$$\ddot{x}_n + 2x_n - x_{n-1} - x_{n+1} = 0, n \neq 0, 1$$
$$\ddot{x}_0 + x_0 - x_{-1} + k(x_0 - x_1) = 0 \tag{17}$$
$$\ddot{x}_1 + x_1 - x_2 + k(x_1 - x_0) = 0$$

Substituting ansatz (12) into equations (17) with $n_L = 0$ and $n_R = 1$ and taking into account the dispersion relation $\omega(q) = 2\sin(q/2)$, after some algebraic manipulations one obtains the following expressions for the transmission and reflection coefficients:

$$\alpha(q) = \frac{(1-k)(1-\exp(-iq))}{2k-1+\exp(iq)}; \beta(q) = \frac{k(1+\exp(iq))}{2k-1+\exp(iq)} \tag{18}$$

Simple manipulations with account of the dispersion relation lead to the formulas:

$$|\alpha(\omega)|^2 = \frac{(1-k)^2 \omega^2}{4k^2 - (2k-1)\omega^2}; |\beta(\omega)|^2 = \frac{k^2(4-\omega^2)}{4k^2 - (2k-1)\omega^2} \tag{19}$$

Computation of the distribution function $f(\omega)$ is complicated problem, since its shape is governed both by applied thermostats and by the introduced defect itself. To obtain crude estimation of the Kapitza resistance, we *assume the equal distribution* of energy in the incident wave, i.e. assume $f(\omega) = 1$ in the propagation zone of the chain. Then, taking into account $v_{gr}(\omega) = \cos\left(\frac{q}{2}\right) = \sqrt{1 - \frac{\omega^2}{4}}$, the integrals in Equation (16) are expressed in the following form:

$$I_1 = \int_0^2 \frac{(1-k)^2 \omega^2}{4k^2 - (2k-1)\omega^2} d\omega = \begin{cases} \frac{k(1-k)^2}{(2k-1)^{\frac{3}{2}}} \left( \ln\left(\frac{k+\sqrt{2k-1}}{k-\sqrt{2k-1}}\right) - \frac{2\sqrt{2k-1}}{k} \right), k > \frac{1}{2} \\ \frac{2k(1-k)^2}{(1-2k)^{\frac{3}{2}}} \left( \frac{\sqrt{1-2k}}{k} - \arctan\left(\frac{\sqrt{1-2k}}{k}\right) \right), \frac{1}{2} > k > 0 \end{cases} \tag{20}$$

$$I_2 = \int_0^2 \frac{k^2(4-\omega^2)\sqrt{1-\frac{\omega^2}{4}}}{4k^2 - (2k-1)\omega^2} d\omega = \begin{cases} \frac{\pi k(3k-2)}{2(2k-1)^2}, k > 1 \\ \frac{\pi k(2-k)}{2}, 0 < k < 1 \end{cases} \tag{21}$$

Final estimation for the Kapitza resistance reads:

$$R_K = \frac{2I_1}{I_2} \tag{22}$$

It is interesting to mention that Expression (23) predicts the non-analytic asymptotic behavior,

$$R_K \sim (k-1)^2 \ln|k-1|, k \to 1 \tag{23}$$

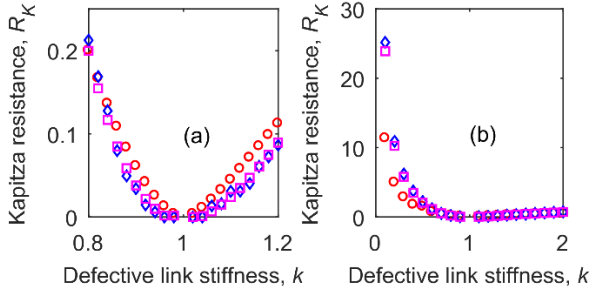

Fig. 4. Comparison of analytical prediction (circle) with molecular dynamics simulations (diamond) and RLL method (square). Here, (a) shows the high resolution plot and (b) shows the low resolution plot. The best fit obtained at $\gamma = 1.3, N_\pm = 10$.

One observes reasonable coincidence, especially in the vicinity of the minimum (Fig. 4). It is not possible to expect perfect coincidence, since the derivation uses an approximation for the spectral density of the energy in the incoming wave. The simulation results demonstrate understandable sensitivity of the Kapitza resistance to particular energy distribution in the chain. Linear chains do not possess any mechanism for the energy redistribution between the modes, therefore these characteristics are completely governed by the chain structure and the thermostats. These findings further corroborate the dependence of the Kapitza resistance on the number and parameters of the thermostats, as presented in Figure 3. To conclude, the linear chain with defect demonstrates well-articulated Kapitza step, but the value of the resistance depends on external factors, and does not characterize the model itself.

Thus, in order to simulate the Kapitza resistance depending only on the structure and temperature, one must include the anharmonicity into the model. This can be done in different ways. The results for basic benchmark models of the nonlinear chains are presented in the next Section.

## 4. Nonlinear chain models
### 4.1. FPU model

In this section we will investigate the Kapitza resistance behavior in chains with β-FPU interaction (3); the nonlinearity is set to $\beta = 0.1$ without reducing the generality. First, we studied the dependence of the Kapitza resistance on the strength of the isotopic defect. As evident from Fig. 5, the Kapitza resistance monotonously increases on either side as the defect becomes lighter or heavier.

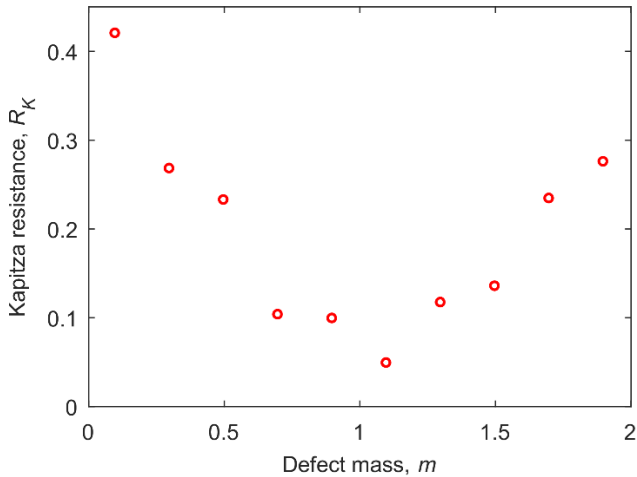

FIG. 5. Variation of the Kapitza resistance $R_K$ with the defect mass $m$ for the chain with β-FPU interaction ($T_+ = 96, T_- = 80, N = 501, N_\pm = 1, \gamma = 1, \beta = 0.1$).

Next, let us consider the dependence of Kapitza resistance on chain length $N$. The simulations are conducted for different chain lengths, $N = 21 - 501$. Fig. 6 reveals that Kapitza resistance vanishes with increasing chain length, and obeys power law variation, $R_K \propto N^{-h}$. The similar relation can be found for thermal conductivity in a pure chain with similar interaction potentials. Even though the reported thermal conductivity divergence exponent varies with different studies, all the reported values were closer to $1/3$. Fortunately, the study [31] shows, at the thermostat temperatures $T_+ = 96$ and $T_- = 80$, the scaling is according to $\kappa(N) = N^{0.37}$, even for very long chain lengths. We chose the same thermostat temperature conditions to explore the Kapitza resistance variation with chain length $N$. The simulation results demonstrates the divergence exponent $h$ is dependent on the isotopic defect mass for the Kapitza resistance. However, it is still closer to the value $1/3$ (see Fig. 6).

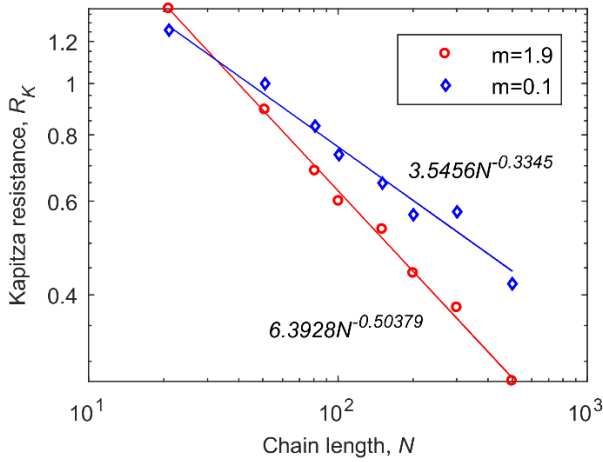

FIG. 6. Power law decrease of the Kapitza resistance $R_K$ with the chain length $N$ for the chain with β-FPU interaction ($T_+ = 96, T_- = 80, N = 501, N_\pm = 1, \gamma = 1, \beta = 0.1$).

In Fig. 7. we show the Kapitza resistance dependence on the coupling friction $\gamma$ (a) and on the number of thermostats $N_\pm$ (b). Since the FPU model demonstrates vanishing Kapitza resistance at thermodynamic limit, we can expect that the Kapitza resistance will depend on the thermostat characteristics. The simulation shows that the Kapitza resistance indeed depends on the thermostat parameters also in β-FPU model. Similar to the linear case, the Kapitza resistance initially varies with number of thermostats, then saturates. The coupling friction and the number of thermostats are taken in the interval $\gamma = 0.1 - 3$ and $N_\pm = 1 - 25$ respectively.

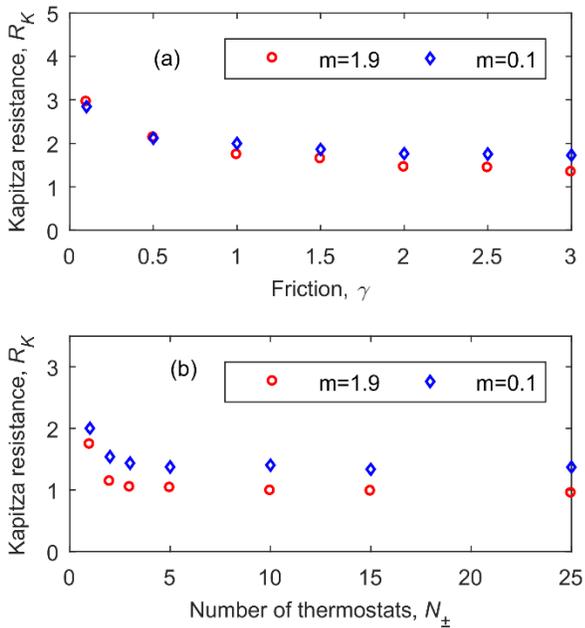

FIG. 7. Dependence of the Kapitza resistance $R_K$ on the coupling friction $\gamma$ (a) and on the number of thermostats $N_\pm$ (b) for the chain with β-FPU interaction ($T_+ = 1.1, T_- = 0.9, N = 501, \beta = 0.1$).

Next, we will consider the Kapitza resistance variation with the temperature. The temperature dependence in the linear model (previous section) as well as in the FPU model are important for the next sections since it stands as a datum to capture various mechanism happening in the following models at very low and high temperatures. Here the temperatures are taken in the interval $T = 0.001 - 10$.

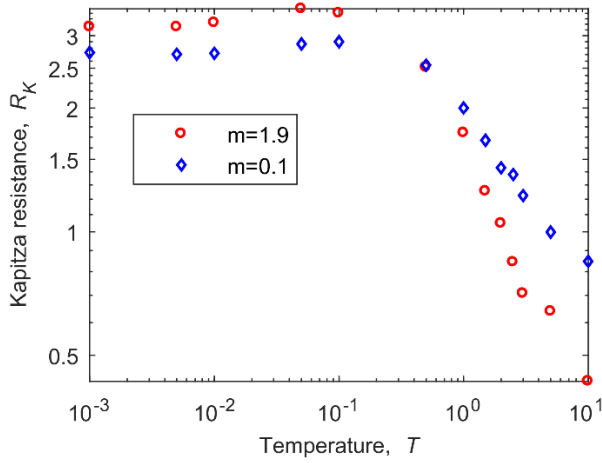

FIG. 8. Dependence of the Kapitza resistance $R_K$ on the chain average temperature $T$ for the chain with β-FPU interaction ($N = 501, N_\pm = 1, \gamma = 1, \beta = 0.1$).

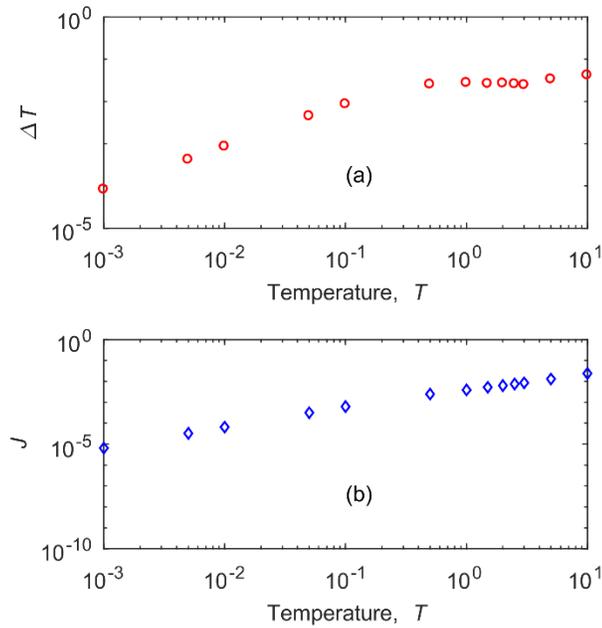

FIG. 9. Kapitza step $\Delta T$ (a) and the heat flux $J$ (b) versus chain average temperature for β-FPU chain with single isotopic defect of $m = 1.9$ ($N = 501, N_\pm = 1, \gamma = 1, \beta = 0.1$).

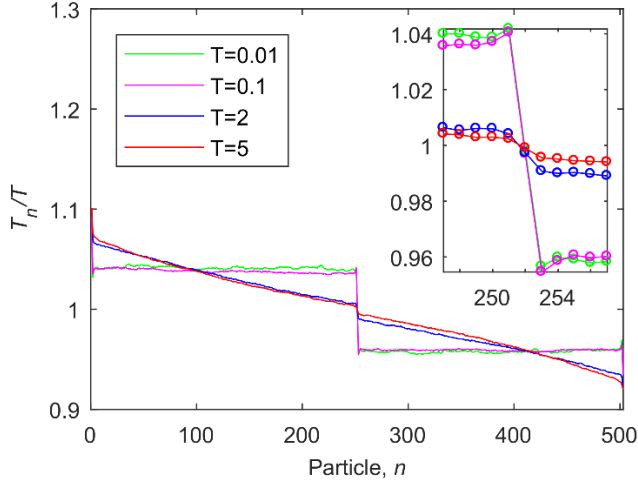

FIG. 10. Normalized temperature profile for different chain average temperatures in β-FPU chain with single isotopic defect of $m = 1.9$ ($N = 501, N_\pm = 1, \gamma = 1, \beta = 0.1$). Inset shows temperature profile near to the defect.

One observes that at low temperatures the simulated Kapitza resistance is almost constant, similar to the linear case. For higher temperatures, the nonlinearity reveals itself and temperature dependence appears. The crossover temperature $T \approx 0.1$ is consistent by order of magnitude with the value of β used in the simulations.

In Fig. 9, we separately plotted the temperature drop $\Delta T$ and the heat flux $J$ versus the chain average temperature $T$. The temperature drop $\Delta T$ has different variation trend after the temperature $T \approx 0.1$ and it makes the temperature dependence of Kapitza resistance in Fig. 8. Fig. 10 shows the normalized temperature profile plotted for different chain average temperature.

One can observe that β-FPU model also does not demonstrate the Kapitza resistance governed solely by local structure of the defect zone. One observes substantial dependence both on the total chain length and the thermostat properties. This finding is apparently related to the well-known divergence of the heat conduction coefficient in this model.

### 4.2. Chain of rotators

This subsection is devoted to the exploration of the chain of rotators with potential function (4). Fig. 11 shows the typical heat flux (a) and temperature (b) profiles in the chain of rotators with isotopic defect (the temperature drop can be found at the defect location).

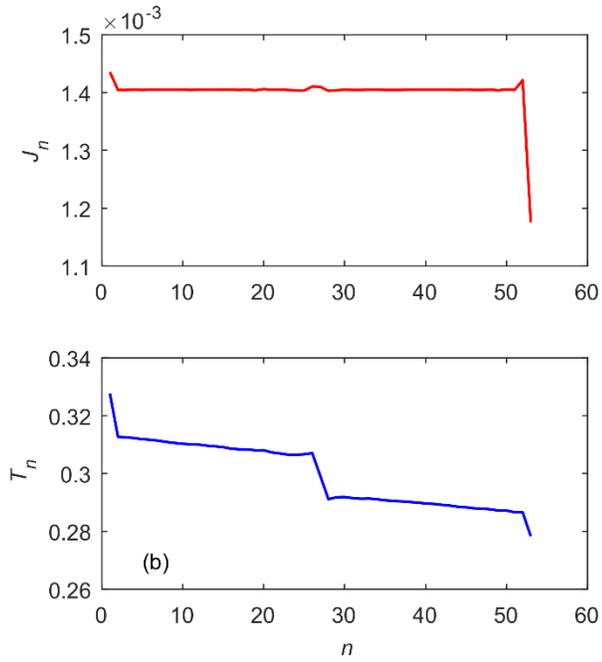

FIG. 11. (a) Average heat flux $J_n$ and (b) local temperature $T_n$ in the chain of rotators for $N = 51, T_+ = 0.33, T_- = 0.27, N_\pm = 1, \gamma = 1$. The isotopic defect mass, $m = 1.9$.

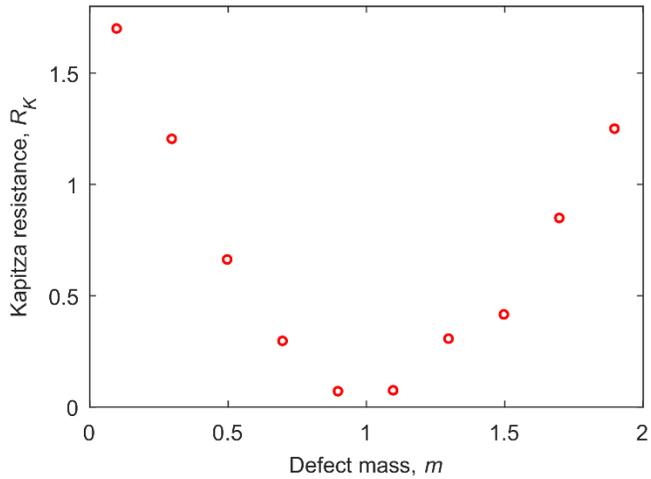

FIG. 12. Variation of the Kapitza resistance $R_K$ with the defect mass $m$ for the chain of rotators ($T_+ = 0.33, T_- = 0.27, N = 501, N_\pm = 1, \gamma = 1$).

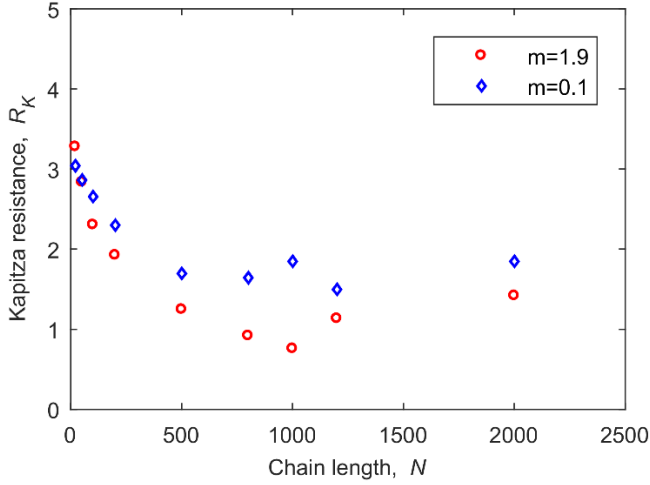

FIG. 13. Convergence of the Kapitza resistance $R_K$ in the chain of rotators with chain length $N$ in the chain of rotators ($T_+ = 0.33, T_- = 0.27, N_\pm = 1, \gamma = 1$).

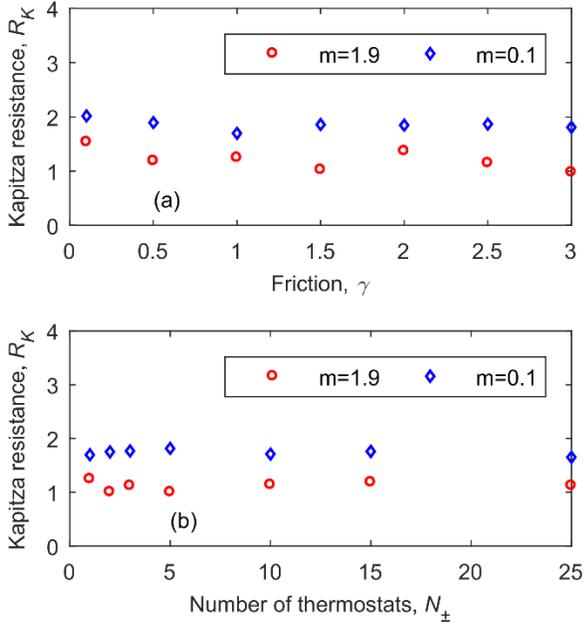

FIG. 14. The Kapitza resistance dependence on the coupling friction $\gamma$ (a) and on number of thermostats $N_\pm$ (b), in the chain of rotators ($N = 501, T_+ = 0.33, T_- = 0.27$)

As shown in Fig. 12, the Kapitza resistance variation with isotopic defect mass follows the similar trend as in the FPU model. The chain of rotators is well-known example for momentum conserving systems with normal heat conductivity. The results presented in Fig. 13 suggest that the Kapitza resistance converges with the chain length ($N$). The chain lengths are selected in the interval $N = 21 - 2001$. In order to further corroborate the convergence, we have additionally studied the Kapitza resistance dependence on the thermostat coupling friction $\gamma$ and the number of thermostats $N_\pm$. Fig. 14(a) presents the Kapitza resistance versus friction and Fig. 14(b) - the Kapitza resistance versus number of thermostats. From Fig. 14, it is obvious that the Kapitza resistance is independent of both thermostat characteristics. Thus, in the chain of rotators the Kapitza resistance is indeed a local property of the defect.

Figure 15 presents the dependence of the Kapitza resistance on the average temperature in the chain. Most interesting feature of this dependence is a dip in the resistance for temperature interval $T \approx 0.2 - 0.3$. It is possible to attribute this dip to the beginning of intense activation of the rotobreathers in this temperature interval [21]; then, the chain becomes strongly non-homogeneous and the presence of the single isolated isotopic defect only slightly affects the heat flux. The temperature effect is explored only up to $T = 0.6$, since after that the Kapitza step (temperature drop at the defect) becomes very small and comparable with the noise levels in the simulation.

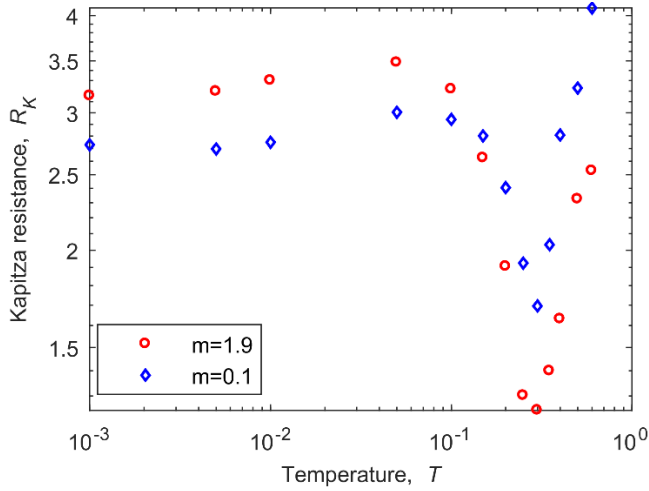

FIG. 15. Temperature dependence of the Kapitza resistance in the chain of rotators ($N = 501$, $\gamma = 1$, $N_\pm = 1$).

### 4.3. Frenkel – Kontorova model

Finally, we address the Kapitza resistance in Frenkel-Kontorova model (5). The dependence on the mass of the isotopic defect strength shows more similarity to the harmonic model, especially for very lighter and heavier defects (see Fig. 16).

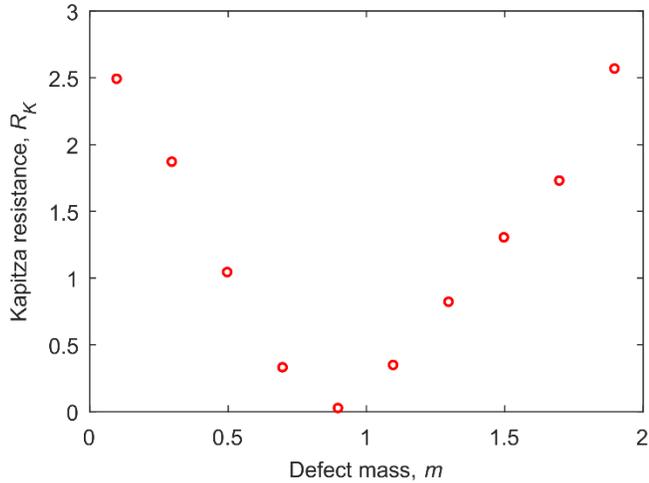

FIG. 16. Variation of the Kapitza resistance $R_K$ with the defect mass $m$ in the Frenkel-Kontorova model ($T_+ = 3.3$, $T_- = 2.7$, $N = 501$, $N_\pm = 1$, $\gamma = 1$).

The Frenkel-Kontorova model is a nonlinear lattice with on-site potential; in such systems, the convergence of the heat conduction coefficient is commonly adopted as well-established [23]. Figure 17 demonstrates the convergence of the Kapitza resistance in Frenkel-Kontorova model at $T = 3$. The chain length is taken in the interval $N = 21 - 1001$. The convergent behavior is further corroborated by the stable Kapitza resistance that is independent of the thermostat coupling friction $\gamma$ (Fig. 18(a)) and the number of thermostats $N_\pm$ (Fig. 18(b)).

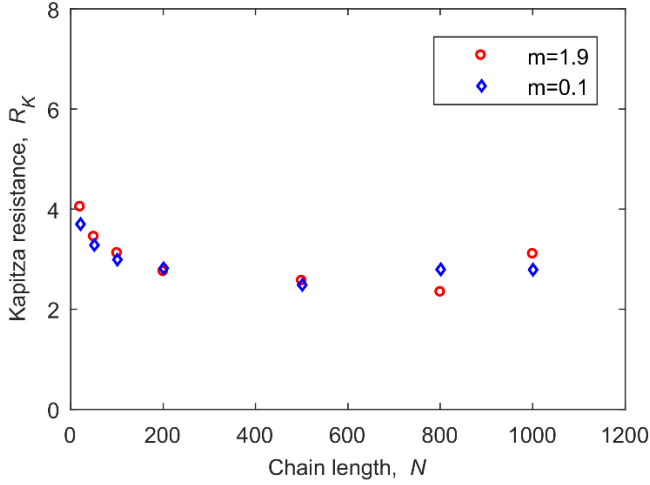

FIG. 17. Convergence of the Kapitza resistance in the Frenkel-Kontorova model with chain length $N$ in the Frenkel-Kontorova model ($T_+ = 3.3, T_- = 2.7, N_\pm = 1, \gamma = 1$).

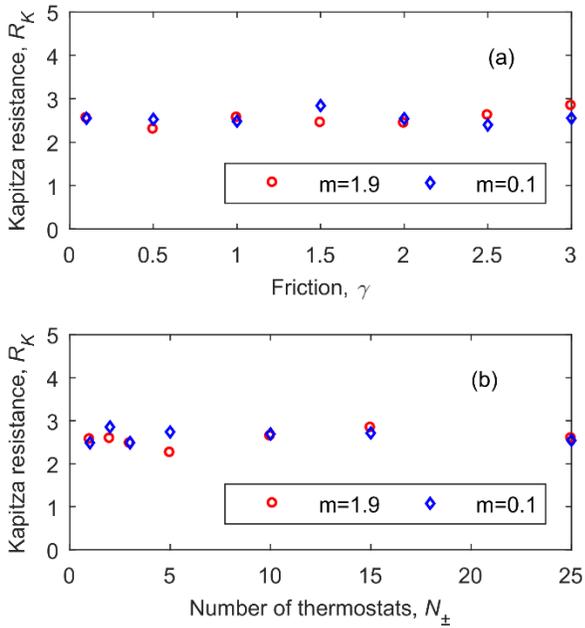

FIG. 18. The Kapitza resistance dependence on the coupling friction $\gamma$ (a) and on the number of thermostats $N_\pm$ (b), in the Frenkel-Kontorova model ($N = 501, T_+ = 3.3, T_- = 2.7$).

Dependence of the Kapitza resistance on the average temperature in the Frenkel-Kontorova model is plotted in Figure 19 ($T = 0.01 - 1000$). For low and high temperatures, one observes the constant limits. It nicely conforms to linear limits of Klein-Gordon chain at low temperatures and detached linear chain at high temperatures – in both cases, one expects almost temperature – independent Kapitza resistance. For the intermediate temperature range, one observes the decrease, related to substantially nonlinear behavior of the model. This result also conforms to well-known minimum of the heat conductivity in this model [23], related to governing role of topological solitons in the heat transport in this temperature range.

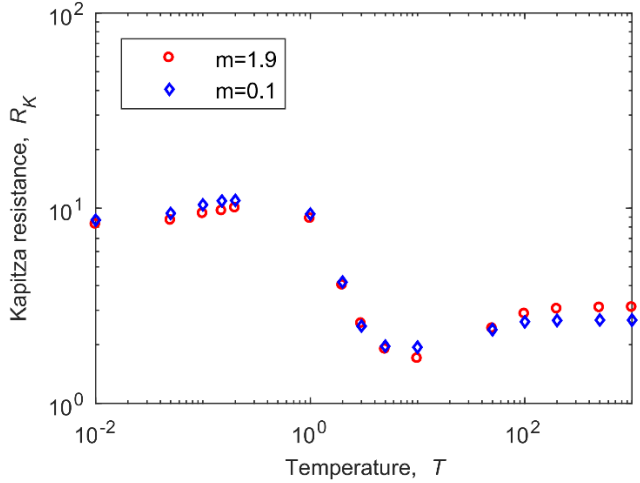

FIG. 19. Dependence of the Kapitza resistance $R_K$ on the chain average temperature $T$ in the Frenkel-Kontorova model ($N = 501$, $N_\pm = 1$, $\gamma = 1$).

## 5. Conclusions

The results presented above allow to conjecture that the division of one-dimensional models into the universality classes established for the bulk one-dimensional heat conduction is valid in certain sense also for the Kapitza resistance. For linear models, this resistance is well-defined and size –independent (contrary to the bulk conductivity), but depends on parameters of the thermostats. Addition of the cubic coupling nonlinearity adds size – and temperature-dependence. In the models with convergent bulk conduction coefficient (chain of rotators and Frenkel – Kontorova model) the Kapitza resistance is also normal, i.e. size – and thermostat – independent.

One can also conclude that the nonlinearity plays double qualitative role in the boundary conduction problem. First of all, it is required to "normalize" the heat flux in the bulk asides the boundary, and to provide the heat flux independent on the system size and on the particular parameters and number of the thermostats. Besides, it can play a substantial role in the processes of boundary phonon scattering. This contribution is readily revealed, since, contrary to the bulk heat conductivity, the Kapitza resistance in the linear chain is finite and temperature - independent. Thus, the temperature dependence of the Kapitza resistance appears solely due to the nonlinearity of the interaction potentials, and points on its role in the boundary scattering in different parametric regimes.